\documentclass[
 amsmath,amssymb,
 aps,
]{revtex4-1}

\usepackage{graphicx}% Include figure files
\usepackage{dcolumn}% Align table columns on decimal point
\usepackage{bm}% bold math
\usepackage{color}

\begin{document}

\title{Microscopic states of Kerr black holes from boundary-bulk correspondence}
\author{Jingbo Wang}
\email{ shuijing@mail.bnu.edu.cn}
\affiliation{Institute for Gravitation and Astrophysics, College of Physics and Electronic Engineering, Xinyang Normal University, Xinyang, 464000, P. R. China}
 \date{\today}
\begin{abstract}
It was claimed by the author that black holes can be considered as topological insulators. They both have boundary modes and those boundary modes can be described by an effective BF theory. In this paper, we analyze the boundary modes on the horizon of black holes with the methods developed for topological insulator. Firstly the BTZ black hole is analysed, and the results are compatible with the previous works. Then we generalize those results to Kerr black holes. Some new results are obtained: dimensionless right- and left-temperature can be defined and have well behaviors both in Schwarzschild limit $a\rightarrow 0$ and extremal limit $a\rightarrow M$. Upon the Kerr/CFT correspondence, we can associate a central charge $c=12 M r_+$ with an arbitrary Kerr black hole if a dual CFT exists. We can identify the microstates of the Kerr black hole with the quantum states of this scalar field. From this identification we can count the number of microstates of the Kerr black hole and give the Bekenstein-Hawking area law for the entropy.
\end{abstract}
\maketitle
\section{Introduction}
It has been well accepted that black holes behave like thermodynamics objects. Due to Bekenstein \cite{bk1} and Hawking \cite{hawk1}, the black holes have temperature and entropy
\begin{equation}\label{1a}
  T_H=\frac{\hbar \kappa}{2\pi},\quad S=\frac{A}{4 G \hbar},
\end{equation}
where $\kappa$ is the surface gravity, and $A$ the area of the horizon. Understanding those properties is a fundamental challenge of quantum gravity, especially identifying the black hole microstates which account for the entropy.

Since this is a difficult problem, it is helpful to map this problem to a solved one. In the previous works \cite{wangti1,wangti2}, the author claimed that the black holes can be considered as kind of topological insulators. For BTZ black hole in three dimensional spacetime this claim is tested in Ref.\cite{wangbms1,wangbms2,wangplb1}. The boundary modes on the horizon of BTZ black hole can be described by two chiral massless scalar fields with opposite chirality. This is the same as topological insulators in three dimensional spacetime (also called quantum spin Hall states) \cite{tibf1}. From those chiral scalar fields one can construct the $W_{1+\infty}$ algebra which contains the near horizon symmetry algebra of BTZ black hole as a subalgebra. The $W_{1+\infty}$ algebra was used to classify the BTZ black holes and give the `W-hairs' of black holes, which maybe essential to solve the information paradox \cite{info1,info2,info3}. The microstates of BTZ black hole are identified as the quantum states of those chiral scalar fields.

For higher dimensional black holes, such as Kerr black holes in four dimension, the boundary modes can be described by boundary BF theory, which is also the same as higher dimensional topological insulators. From the BF theory, one can construct a free scalar field theory. An essential property of topological insulators is the boundary-bulk correspondence \cite{bounbulk1,bounbulk2}, which relates the topological structure of the bulk states to the presence of gapless boundary modes. From those boundary modes one can get some key properties of the bulk states \cite{boson1,boson2}. Just like the BTZ black hole, we can also identify the quantum states of the scalar field with the microstates of Kerr black holes.

In the previous works \cite{wmz,hw1}, it was shown that the degrees of freedom on the horizon can be described by BF theory with sources. This result hold for wide cases, including general relativity in arbitrary dimension \cite{wang1,wh1,wh3}, Lovelock theory \cite{wh2}. Those results are based on the symplectic structure analysis in the framework of loop quantum gravity \cite{rov1,thie1,ash0,ma1}. We start form this BF theory. Firstly we analyze the BTZ black holes. Then we apply the same method to Kerr black holes. For Kerr black holes, we get some new results:
\begin{itemize}
  \item We can define dimensionless  left- and right-temperatures
 $T_{L/R}=\frac{r_+\pm a}{4\pi r_+}$ for all Kerr black holes (including Schwarzschild black hole), which coincide with the temperatures in Kerr/CFT correspondence \cite{kerrcft1,kerrcft2,kerrcft3} in the extremal limit.
  \item For arbitrary Kerr black hole, if a dual conformal field theory (CFT) exist, one can associate a central charge $c=12 M r_+$. It coincides with that in the extremal case for Kerr/CFT correspondence.
  \item  We identify the microscopic states of Kerr black holes with the quantum states of the boundary scalar field and count the number of those states. It can give the Bekenstein-Hawking area law.
\end{itemize}

The paper is organized as follows. In section II, we analyze the BTZ black hole. In section III, the Kerr black hole is analysed with the same method. Section IV is the conclusion. In the following we set $G=\hbar=c=1$.
\section{The BTZ black hole}
In this section, we consider the boundary modes on the horizon of BTZ black hole from the boundary BF theory \cite{wangti2}. To quantize the boundary BF theory, we mainly follow the methods in Ref.\cite{boson1,boson2}.

The metric of BTZ black hole is
\begin{equation}\label{1}
    ds^2=-N^2 dv^2+2 dv dr+r^2 (d\varphi+N^\varphi dv)^2,
\end{equation}
where $N^2=-8 M+\frac{r^2}{L^2}+\frac{16 J^2}{r^2}, N^\varphi=-\frac{4 J}{r^2}$.

Choose the following Newman-Penrose null co-triads
\begin{equation}\label{2}
    l=-\frac{1}{2}N^2 dv+dr,\quad n=-dv,\quad m=r N^\varphi dv+r d\varphi,
\end{equation}
then the corresponding spin connection which is used later is $A_2=\alpha m -\kappa n$ with $\alpha=N^\varphi,\kappa=r/L^2-r(N^\varphi)^2$.

For later use, we define new coordinate $dv'=dv/\gamma$ with the coefficient $\gamma$ determined later. Then on the horizon $\Delta:r=r_+$,
\begin{equation}\label{2a}
  m=(r N^\varphi dv+r d\varphi)|_\Delta=-\frac{\gamma r_- }{L}dv'+r_+ d\varphi.
\end{equation}

\subsection{The canonical formula}
On the horizon $r=r_+$, we choose the coordinates $(x^0,x^1)=(v',\varphi)$. The boundary BF theory on the horizon $\Delta$ is given by \cite{wangti2}
\begin{equation}\label{3}
    S=\int_{\Delta}BF=\int_{\Delta}B {\rm d}A,
\end{equation}
with the constraints
\begin{equation}\label{4}
    {\rm d}B=\frac{1}{8\pi}m,\quad {\rm d} A=0,
\end{equation}
where $A$ is the non-rotating component of the spin connection $A_2$.

The canonical form is
\begin{equation}\label{6}\begin{split}
    S=\int_{\Delta}dv' d \varphi B(\partial_0 A_1-\partial_1 A_0)
    =\int_{\Delta}dv' d \varphi (B\partial_0 A_1+A_0\partial_1 B).
\end{split}\end{equation}
Choosing the gauge $A_0=0$ gives $A_1=\partial_1 \phi$, so
\begin{equation}\label{8}\begin{split}
    S=-\int_{\Delta}dv' d \varphi \partial_1 B\partial_0 \phi.
\end{split}\end{equation}

Since the horizon is a null hypersurface, the metric is degenerate. To make things easier, we assume that the effective metric of the horizon is
\begin{equation}\label{9}
  \tilde{ds}^2=-dv'^2+r_+^2 d\varphi^2,
\end{equation}
then
\begin{equation}\label{10}
  \sqrt{-g}=r_+, \quad \epsilon^{01}=\frac{1}{r_+}.
\end{equation}
The action can be rewritten as
\begin{equation}\label{11}\begin{split}
    S=-\int_{\Delta}d^2 x\sqrt{-g} \frac{1}{\sqrt{-g}}\partial_1 B\partial_0 \phi
    =\int_{\Delta}d^2 x\sqrt{-g} \pi \dot{\phi},
\end{split}\end{equation}
where $\pi=-\frac{1}{\sqrt{-g}}\partial_1 B$ is the canonical momentum. It is easy to show that the Hamiltonian is zero. To describe a relativistic dynamics, one can add a Hamiltonian to get \cite{tibf1,boson2}
\begin{equation}\label{12}\begin{split}
    S'=\int_{\Delta}d^2 x\sqrt{-g} (\pi \dot{\phi}-\mathcal{H}(\pi,\phi))=\int_{\Delta}d^2 x\sqrt{-g} (\pi \dot{\phi}-\frac{1}{2m_0}\pi^2-\frac{m_0}{2}g^{ij}\partial_i \phi \partial_j \phi),
\end{split}\end{equation}
where $m_0$ is a free parameter. It is the simplest choice which can give the non-trivial dynamics on the boundary. Other forms of Hamiltonian can also be chosen, just showed in Ref.\cite{boson2}.

The equations of motion are
\begin{equation}\label{13}
  \pi=m_0\dot{\phi},\quad \dot{\pi}=m_0\Delta \phi,
\end{equation}
where $\Delta$ is the Laplace-Beltrami operator.

The above equation can be recast into a duality relation
\begin{equation}\label{14}
  \epsilon^{\mu\nu}\partial_\nu B=m_0 \partial^\mu \phi.
\end{equation}

The action (\ref{12}) is just the action for massless scalar field
\begin{equation}\label{15}\begin{split}
    S'=\frac{m_0}{2}\int_{\Delta}d^2 x\sqrt{-g}g^{\mu\nu}\partial_\mu \phi \partial_\nu \phi.
\end{split}\end{equation}

The Hamiltonian can be given by
\begin{equation}\label{15a}\begin{split}
  H=\frac{m_0}{2}\oint d\varphi\sqrt{-g} ((\partial_0 \phi)^2+ (\frac{\partial_1 \phi}{r_+})^2).
 \end{split}\end{equation}

We can also define the angular momentum
\begin{equation}\label{15b}\begin{split}
 J=m_0\oint d\varphi\sqrt{-g} (\partial_0 \phi \partial_1 \phi)
 \end{split}\end{equation}
\subsection{Quantization}
In this section we quantize this massless scalar field (\ref{15}). Expanding the fields with the Fourier modes gives
\begin{equation}\label{16}\begin{split}
  \phi(v',\varphi)=\phi_0+p_v v'+p_\varphi \varphi+ \sqrt{\frac{1}{m_0 A}}\sum_{n\neq 0}\sqrt{\frac{1}{2 \omega_n}}[a_n e^{-i(\omega_n v'-k_n \varphi)}+a^+_n e^{i(\omega_n v'-k_n \varphi)}],\\
  B(v',\varphi)=-m_0(B_0+\frac{p_\varphi}{r_+}v'+r_+ p_v \varphi)+\sqrt{\frac{4 m_0}{r_+^2 A}}\sum_{n\neq 0}\sqrt{\frac{1}{(2 \omega_n)^3}}k_n[a_n e^{-i(\omega_n v'-k_n \varphi)}+a^+_n e^{i(\omega_n v'-k_n \varphi)}],
\end{split}\end{equation}
where $\omega_n=\frac{|n|}{r_+},k_n=n$ and $A=2\pi r_+$ is the length of the circle. It is straight to show that the above expressions satisfy the dual relation (\ref{14}).

The quantum field operators satisfy the commutative relation
\begin{equation}\label{17}
  [\hat{\phi}(v',\varphi), \hat{\pi}(v',\varphi')]=i \delta(\varphi-\varphi'),
\end{equation}
which gives
\begin{equation}\label{18}
  [\hat{\phi}_0,m_0 \hat{p}_v]=\frac{i}{A},\quad [\hat{a}_n, \hat{a}^+_m]=\delta_{n,m}.
\end{equation}

We can also consider $B$ and $\partial_1 \phi$ as two canonical variables, and lead to a further commutation relation
\begin{equation}\label{19}
 [\hat{B}_0,\frac{m_0}{r_+} \hat{p}_\varphi]=-\frac{i}{A}.
\end{equation}

Since the zero mode $\phi_0,B_0$ are constants on the cylinder, the spectrum of the canonical momentum are quantized according to
\begin{equation}\label{20}
 m_0 p_v= \frac{n_1}{A},\quad \frac{m_0}{r_+} p_\varphi=\frac{n_2}{A}\quad n_1,n_2 \in Z.
\end{equation}

For the BTZ black hole, the $B$ field satisfy the constraint (\ref{4}), or with the component
\begin{equation}\label{21}
  \partial_0 B=-\frac{r_- \gamma}{8\pi L},\quad \partial_1 B=\frac{r_+}{8\pi}.
\end{equation}

Submitting the expression (\ref{16}) gives
\begin{equation}\label{23}
 p_v= -\frac{1}{8\pi m_0},\quad p_\varphi=\frac{\gamma r_+ r_- }{8\pi L m_0}.
\end{equation}
Combing with the Equ.(\ref{20}) give the quantization condition
\begin{equation}\label{24}
  r_+=4 n_1,\quad r_-=4 n_2 \frac{L}{\gamma r_+}, \quad n_1,n_2 \in N.
\end{equation}
The dimensionless right- and left-temperature are defined as
\begin{equation}\label{29}
  T_{R/L}=\frac{r_+\pm r_-}{2\pi L}.
\end{equation}

To fix the coefficient $\gamma$ we make the following assumption,
\begin{equation}\label{29a}
  T_{R/L}\propto p_v\mp \frac{p_\varphi}{r_+},
\end{equation}
which gives $\gamma=\frac{L}{r_+}$. So
\begin{equation}\label{29c}
  r_+=4 n_1 ,\quad r_-=4 n_2 , \quad n_1,n_2 \in N.
\end{equation}
From the above expression, the length $L_+,L_-$ of the outer and inner horizon of black hole are also quantized according to
\begin{equation}\label{29d}
  L_+=2\pi r_+=8\pi n_1,\quad L_-=2 \pi r_-=8\pi n_2, \quad n_1,n_2 \in N.
\end{equation}

The quantum version of Hamiltonian (\ref{15a}) and angular momentum (\ref{15b}) are given by
\begin{equation}\label{22a}\begin{split}
  \hat{H}=\pi m_0 r_+(\hat{p}_v^2+\frac{\hat{p}^2_\varphi}{r_+^2})+\sum_{n\neq 0} \frac{|n|}{r_+}\hat{a}_n^+ \hat{a}_n,\\
   \hat{J}=2 \pi m_0 r_+ \hat{p}_v \hat{p}_\varphi +\sum_{n\neq 0} n \hat{a}_n^+ \hat{a}_n,
\end{split}\end{equation}
where we omit the zero-point energy.

To fix the parameter $m_0$, we assume
\begin{equation}\label{22c}
2 \pi m_0 r_+ p_v p_\varphi=J=\frac{r_+ r_-}{4 L},
\end{equation}
which gives
\begin{equation}\label{22d}
  m_0=\frac{L}{8 \pi}.
\end{equation}

So the dimensionless temperatures can be expressed by the zero-mode part as
\begin{equation}\label{26}
  T_{R/L}=-\frac{r_+}{2\pi}(p_v\mp \frac{p_\varphi}{r_+}),
\end{equation}
and satisfy
\begin{equation}\label{26a}
  \frac{2}{T_H}=\frac{\gamma r_+}{T_R}+\frac{\gamma r_+}{T_L},
\end{equation}
where $T_H$ is Hawking temperature for the black hole.

The scalar field $\phi(v',\varphi)$ can be considered as collectives of harmonic oscillators, and a general quantum state can be represented as $|p_v,p_\varphi;\{n_k\}>\sim (a^+_1)^{n_1}\cdots (a^+_k)^{n_k}|p_v,p_\varphi>$ where $p_v,p_\varphi$ are zero mode parts, and $\{n_k\}$ are oscillating parts. As was shown in Ref.\cite{wangbms2,wangplb1}, the BTZ black hole ground state corresponds to the zero mode part, thus
\begin{equation}\label{25c}
  <p_v,p_\varphi;\{0\}|\hat{J}|p_v,p_\varphi;\{0\}>=J,\quad <p_v,p_\varphi;\{0\}|\hat{H}|p_v,p_\varphi;\{0\}>=\gamma M=M L/r_+.
\end{equation}
The parameter $\gamma$ appears because the energy $M$ is associated with the time coordinate $v$, and for $v'$ one have
\begin{equation}\label{26b}
  M' \sim \frac{\partial}{\partial v'}= \gamma\frac{\partial}{\partial v}\sim \gamma M.
\end{equation}

On the other hand, the microstates of BTZ black hole can be represented by $|0,0;\{n_k\}>$, and satisfy
\begin{equation}\label{25e}
  \frac{1}{c}<0,0;\{n_k\}|\hat{J}|0,0;\{n_k\}>=J,\quad \frac{1}{c}<0,0;\{n_k\}|\hat{H}|0,0;\{n_k\}>=M L/r_+,
\end{equation}
where $c=3L/2G$ is the Brown-Henneaux central charge. It can be written as
\begin{equation}\label{25ee}
  \sum_{k\neq 0} k n_k=c J,\quad \sum_{k\neq 0} |k| n_k=c M L, \quad n_k \in N^+.
\end{equation}
The above equations are equivalent to
\begin{equation}\label{25eee}
  \sum_{k> 0} k n^+_k=\frac{c}{2}(M L+J),\quad \sum_{k< 0} (-k) n^-_k=\frac{c}{2}(M L-J), \quad n^{\pm}_k \in N^+.
\end{equation}
Different sequences $\{n_k\}$ correspond to different microstates of the BTZ black hole with fixed $(M,J)$. For non-extremal black holes, the total number of the microstates for BTZ black hole with parameters $(M,J)$ can be calculated through Hardy-Ramanujan formula,
\begin{equation}\label{27}
  p(N)\simeq \frac{1}{4N \sqrt{3}}\exp(2 \pi \sqrt{\frac{N}{6}}).
\end{equation}
The result is
\begin{equation}\label{25f}
  N(M,J) \simeq \frac{1}{c (M L+J) c (M L-J)}\exp(2 \pi \sqrt{c \frac{M L+J}{12}}+2 \pi \sqrt{c \frac{M L-J}{12}}).
\end{equation}
So the entropy of the BTZ black hole is given by
\begin{equation}\label{25g}
  S=\ln N(M,J)=\frac{2 \pi r_+}{4}-2\ln (r_+^2-r_-^2)+\cdots,
\end{equation}
which is just the Bekenstein-Hawking entropy formula with some low order corrections.

For extremal BTZ black hole $J=M L$, the constraints (\ref{25eee}) become
\begin{equation}\label{25h}
  \sum_{k>0}k n^+_k=c M L, \quad n^+_k \in N^+,
\end{equation}
so the entropy is given by
\begin{equation}\label{30a}
  S=\ln N(M)=\ln (\frac{1}{c M L}\exp(2 \pi \sqrt{c \frac{M L}{6}}))=\frac{2 \pi r_+}{4}-2\ln r_++\cdots,
\end{equation}
which is the same as the results in ``horizon fluff" proposal \cite{fluff3}.
\section{Kerr black hole}
In this section we analyse the Kerr black hole with the same method. The metric of Kerr black hole can be written as \cite{kerr1}
\begin{equation}\label{30}
  ds^2=-(1-\frac{2 M r}{\rho^2})dv^2+2 dv dr-2 a \sin^2 \theta dr d\varphi-\frac{4 a M r \sin^2 \theta}{\rho^2}dv d\varphi+\rho^2 d\theta^2+\frac{\Sigma^2 \sin^2 \theta}{\rho^2}d\varphi^2,
\end{equation}
where $\rho^2=r^2+a^2 \cos^2 \theta, \Delta^2=r^2-2 M r+a^2, \Sigma^2=(r^2+a^2)\rho^2+2 a^2 M r \sin^2 \theta$. The horizon is localized at $r=r_+$.

A suitable null co-tetrads $(l,n,m,\bar{m})$ can be chosen as
\begin{equation}\label{31}\begin{split}
l=-\frac{\Delta}{2(r^2+a^2)}dv+\frac{\rho^2}{r^2+a^2}dr+ \frac{\Delta a \sin^2 \theta}{2(r^2+a^2)}d\varphi,\\
n=\frac{r^2+a^2}{\rho^2}(-dv+a \sin^2 \theta d\varphi),\\
m=-\frac{a \sin \theta}{\sqrt{2}\tilde{\rho}}dv+ \frac{(r^2+a^2) \sin \theta}{\sqrt{2}\tilde{\rho}}d\varphi+\frac{i}{\sqrt{2}}\bar{\tilde{\rho}}d\theta,
\end{split}\end{equation}
where $\tilde{\rho}=r+i a \cos \theta$. It is easy to show that area element of the horizon $\Delta$ is
\begin{equation}\label{31a}
  -i m\wedge \bar{m}=a \sin\theta dv\wedge d\theta+(r_+^2+a^2)\sin\theta d\theta\wedge d\varphi=a \gamma \sin\theta dv'\wedge d\theta+r_0^2\sin\theta d\theta\wedge d\varphi,
\end{equation}
with new coordinate $dv'=dv/\gamma$ and $r_0^2=r_+^2+a^2$.
\subsection{The canonical formula}
On the horizon $\Delta: r=r_+$, we choose $(x^0,x^1,x^2)=(v',\theta,\varphi)$. The boundary BF theory on the horizon $\Delta$ is
\begin{equation}\label{33}
    S=\int_{\Delta}BF=\int_{\Delta}B {\rm d}A,
\end{equation}
with the constraints
\begin{equation}\label{34}
    {\rm d}B=-i\frac{1}{8\pi}m\wedge \bar{m},\quad {\rm d} A=0.
\end{equation}

Assume that the effective metric of the horizon is
\begin{equation}\label{39}
  \tilde{ds}^2=-dv'^2+r^2_+(d\theta^2+\sin^2 \theta d\varphi^2),
\end{equation}
then
\begin{equation}\label{40}
  \epsilon_{012}=\sqrt{-g}=r_+^2 \sin\theta, \quad \epsilon^{012}=\epsilon^{12}=\frac{1}{r_+^2 \sin\theta}.
\end{equation}

Following the same methods as in the BTZ black hole case, we choose the gauge $B_0=A_0=0$ to gives $A_i=\partial_i \phi$, and the action becomes
\begin{equation}\label{38}\begin{split}
    S=-\int_{\Delta}d^3 x (\partial_2 B_1-\partial_1 B_2)\partial_0 \phi
     =\int_{\Delta}d^3 x\sqrt{-g} \pi \dot{\phi},
\end{split}\end{equation}
where $\pi=-\epsilon^{ij}\partial_i B_j$ is the canonical momentum. We add a Hamiltonian to get
\begin{equation}\label{42}\begin{split}
    S'=\int_{\Delta}d^3 x\sqrt{-g} (\pi \dot{\phi}-\frac{1}{2m_0}\pi^2-\frac{m_0}{2}g^{ij}\partial_i \phi \partial_j \phi)=\frac{m_0}{2}\int_{\Delta}d^3 x\sqrt{-g}g^{\mu\nu}\partial_\mu \phi \partial_\nu \phi,
\end{split}\end{equation}
where $m_0$ is a mass parameter to adjust the mismatch of dimensions between boundary and bulk. This is just a free scalar field theory.

The equations of motion can be recast into a duality relation
\begin{equation}\label{44}
  \epsilon^{\mu\nu\rho}\partial_\nu B_\rho=m_0 \partial^\mu \phi.
\end{equation}

The Hamiltonian can be written as
\begin{equation}\label{45a}\begin{split}
  H=\frac{m_0}{2}\int_{\Sigma}d^2 x\sqrt{-g}(\partial_0 \phi \partial_0 \phi+g^{ij}\partial_i \phi \partial_j \phi),
  \end{split}\end{equation}
where we omit the zero-point energy.

The angular momentum can be defined by
\begin{equation}\label{45b}\begin{split}
 J=m_0 \int_{\Sigma}d^2 x\sqrt{-g} (\partial_0 \phi \partial_2 \phi).
 \end{split}\end{equation}
\subsection{Quantization}
Next we quantize the massless scalar field (\ref{42}). Expanding the fields with the Fourier modes gives
\begin{equation}\label{46}\begin{split}
  \phi=\phi_0+p_v v'+p_\theta \ln (\cot\frac{\theta}{2})+p_\varphi\varphi+\sqrt{\frac{1}{m_0 A}}\sum_{l> 0}\sum_{m=-l}^{m=l}\sqrt{\frac{1}{2\omega_l}}[a_{l,m} e^{-i \omega_l v'}Y^m_l(\theta,\varphi)+a^+_{l,m} e^{i \omega_l v'}(Y^m_l)^*(\theta,\varphi)],\\
  B_1=m_0(B_{10} r_++\frac{p_\varphi}{\sin\theta} v'+\frac{1}{2}r_+^2 \sin\theta p_v \varphi-2 \sqrt{\frac{1}{m_0 A}}\sum_{l> 0}\sum_{m=-l}^{m=l}\sqrt{\frac{1}{(2\omega_l)^3}}\frac{m}{\sin\theta}[a_{l,m} e^{-i \omega_l v'}Y^m_l(\theta,\varphi)+a^+_{l,m} e^{i \omega_l v'}(Y^m_l)^*(\theta,\varphi)]),\\
   B_2=m_0(B_{20} r_++p_\theta v'-\frac{1}{2}r_+^2 \sin\theta p_v\theta-2 i \sqrt{\frac{1}{m_0 A}}\sum_{l> 0}\sum_{m=-l}^{m=l}\sqrt{\frac{1}{(2\omega_l)^3}}\sin\theta\frac{\partial}{\partial\theta}[a_{l,m} e^{-i \omega_l v'}Y^m_l(\theta,\varphi)-a^+_{l,m} e^{i \omega_l v'}(Y^m_l)^*(\theta,\varphi)]),
\end{split}\end{equation}
where $\omega^2_l=\frac{l(l+1)}{r_+^2}$, $Y^m_l(\theta,\varphi)$ are spherical harmonics and $A=4\pi r_+^2$. It is straight to show that the above expression satisfy the dual relation (\ref{44}).

The commutative relation between $\hat{\phi}$ and $\hat{\pi}$ is
\begin{equation}\label{47}
  [\hat{\phi}(v',\vec{x}), \epsilon^{ij}\partial_i \hat{B}_j(v',\vec{y})]=-i \delta^2(\vec{x}-\vec{y}),
\end{equation}
and gives
\begin{equation}\label{48}
  [\hat{\phi}_0,m_0 \hat{p}_v]=\frac{i}{A},\quad [\hat{a}_{l,m}, \hat{a}^+_{l',m'}]=\delta_{l,l'}\delta_{m,m'}.
\end{equation}

We can also consider $B_i$ and $\epsilon^{ij}\partial_j \phi$ as two canonical variables, and lead to two further commutation relations
\begin{equation}\label{49}
 [\hat{B}_{10},m_0 r_+\frac{ \hat{p}_\varphi}{r_+^2 \sin\theta}]=\frac{i}{A},\quad [\hat{B}_{20},m_0 r_+\frac{\hat{p}_\theta}{r_+^2 } ]=-\frac{i}{A}.
\end{equation}

For the Kerr black hole, the $B$ field satisfy the constraint (\ref{34}), or with the components
\begin{equation}\label{51}
  \partial_0 B_1=\frac{a \gamma \sin\theta}{8\pi},\quad \partial_1 B_2-\partial_2 B_1=\frac{r_0^2\sin\theta}{8\pi},\quad \partial_0 B_2=0,
\end{equation}
which gives
\begin{equation}\label{51a}
  p_v=-\frac{r_0^2}{8 \pi m_0 r_+^2},\quad p_\varphi=\frac{a\sin^2\theta \gamma}{8 \pi m_0},\quad p_\theta=0.
\end{equation}
To get rid the trigonometric functions we define $|\hat{p}_\varphi|=\frac{a \gamma}{8 \pi m_0}$ which can be considered as quantum operator of $p_\varphi$.

Then the commutation relations give
\begin{equation}\label{52}
  [\hat{\phi}_0,-\frac{4\pi \hat{r}_0^2}{8\pi}]=i,\quad [\hat{B}_{10},\frac{\hat{a} r_+ \gamma}{2}]=i.
\end{equation}
Since the zero mode $\phi_0,B_{i0}$ are constant on the horizon, the canonical momentum are quantized with
\begin{equation}\label{53}
\frac{4\pi r_0^2}{8\pi}=n_1,\quad \frac{a r_+ \gamma}{2}=n_2, \quad n_1,n_2 \in N.
\end{equation}

To fix the parameters $\gamma, m_0$, we define the dimensionless left- and right-temperature as (\ref{26})
\begin{equation}\label{59}
  T_{R/L}=-\frac{r_+}{2\pi}(p_v\mp \frac{|p_\varphi|}{r_+})=\frac{r_0^2}{16\pi^2 m_0 r_+}(1\pm \frac{a \gamma r_+}{r_0^2}).
\end{equation}
We require that in the extremal limit $a\rightarrow M$, $T_L\rightarrow 0$. A natural choice should be
\begin{equation}\label{59a}
  \gamma=\frac{r_0^2}{r_+^2}.
\end{equation}
We assume that they also satisfy the condition (\ref{26a}), that is
\begin{equation}\label{59c}
 \frac{2}{T_H}=\frac{\gamma r_+}{T_R}+\frac{\gamma r_+}{T_L}\Rightarrow  m_0=\frac{M}{2\pi}.
\end{equation}

So the dimensionless temperatures are given by
\begin{equation}\label{59b}
  T_{R/L}=\frac{1}{4\pi}(1\pm \frac{a}{r_+}).
\end{equation}
Notice that in the Schwarzschild case $a=0$, those temperature reduce to $T=\frac{1}{4\pi}$. In the extremal limit $a\rightarrow M$, they coincide with the temperatures in Kerr/CFT correspondence $T_{R/L}=\frac{1}{4\pi a}(r_+\pm r_-)$.

If we assume a dual CFT exist for arbitrary Kerr black hole, the central charge $c$ can be obtained through the Cardy formula
\begin{equation}\label{60}
  S_{Cardy}=\frac{\pi^2}{3}c (T_R+T_L)=S_{BH}=2\pi M r_+,
\end{equation}
which gives
\begin{equation}\label{60b}
 c=12 M r_+ =\frac{6 S_H}{\pi}.
\end{equation}

The quantized condition (\ref{53}) gives
\begin{equation}\label{60a}
  A_H=4\pi r_0^2=8\pi n_1,\quad J=n_2, \quad n_1, n_2 \in N,
\end{equation}
that is, the area and the angular momentum of the Kerr black hole are both quantized. The area spectrum has the same form as that in Ref.\cite{bek2,mag1}.

The scalar field $\phi(v',\theta,\varphi)$ can be considered as collectives of harmonic oscillators, and a general state can be represented as $|p_v,p_\varphi;\{n_{l,m}\}>\sim (\hat{a}^+_{1,m})^{n_{1,m}}\cdots (\hat{a}^+_{l,m})^{n_{l,m}}|p_v,p_\varphi>$ where $|p_v,p_\varphi>$ are zero mode part, and $|\{n_{l,m}\}>$ are occupation numbers for the oscillator part. The Hamiltonian operator and the angular momentum operator for the free scalar field can be written as
\begin{equation}\label{55}\begin{split}
  \hat{H}_{free}= \frac{m_0}{2} A(\hat{p}_v^2+\frac{|\hat{p}_\varphi|^2}{r_+^2})+\sum_{l> 0}\sum_{m=-l}^{m=l} \frac{\sqrt{l(l+1)}}{r_+}\hat{a}^+_{l,m} \hat{a}_{l,m},\\
  \hat{J}=m_0 A \hat{p}_v |\hat{p}_\varphi|+\sum_{l> 0}\sum_{m=-l}^{m=l}m \hat{a}^+_{l,m} \hat{a}_{l,m},
 \end{split}\end{equation}
 Unlike the BTZ black hole case, in high dimensional spacetime $D\geq 4$, general relativity have local degrees of freedom. So it is natural to consider the scalar field with interaction
\begin{equation}\label{55a}
  \hat{H}_{full}=\hat{H}_{free}+\hat{H}_{int}.
\end{equation}
The calculation of entropy for higher dimensional Kerr black holes \cite{wangkerr3} suggest to consider the full Hamiltonian has spectrum
\begin{equation}\label{55b}\begin{split}
  \hat{H}_{full}= \frac{m_0}{2} A(\hat{p}_v^2+\frac{|\hat{p}_\varphi|^2}{r_+^2})+\sum_{l>0}\sum_{m=-l}^{m=l} \frac{|m|}{r_+}\hat{a}^+_{l,m} \hat{a}_{l,m}\\=\hat{H}_0+\sum_{l> 0}\sum_{m=-l}^{m=l} \frac{|m|}{r_+}\hat{n}_{l,m}=\hat{H}_0+\sum_m \frac{|m|}{r_+}\hat{n}_m, \quad m\neq 0,
\end{split}\end{equation}
  where $\hat{n}_m=\sum_{l>0} \hat{n}_{l,m}, \hat{n}_{l,m}=\hat{a}^+_{l,m} \hat{a}_{l,m}$ are number operators. The angular momentum operator can also be rewritten as
\begin{equation}\label{55c}\begin{split}
    \hat{J}=\hat{J}_0+\sum_m m \hat{n}_m.
 \end{split}\end{equation}

It is easy to show that the zero mode part takes half value of the angular momentum and the energy for the Kerr black hole, that is
\begin{equation}\label{57}
  <p_v,p_\varphi;\{0\}|\hat{J}|p_v,p_\varphi;\{0\}>=\frac{J}{2}=\frac{a M}{2},\quad <p_v,p_\varphi;\{0\}|\hat{H}|p_v,p_\varphi;\{0\}>=\frac{\gamma}{2}\frac{M}{2}.
\end{equation}
One may confuse with the $\frac{M}{2}$ in expression (\ref{57}). Actually this can be explained by associating the Hamiltonian with the enthalpy \cite{komar2} or thermodynamic potential \cite{highd5,wangkerr3} of the black hole.

On the other hand, we can define the microscopic states of the Kerr black hole as states represented by $|0,0;\{n_m\}>$. They take another half value of the angular momentum and the energy for the Kerr black hole, that is,
\begin{equation}\label{61}
  \frac{1}{c}<0,0;\{n_m\}|\hat{J}|0,0;\{n_m\}>=\frac{J}{2},\quad \frac{1}{c}<0,0;\{n_m\}|\hat{H}|0,0;\{n_m\}>=\frac{\gamma}{2}\frac{M}{2},
\end{equation}
where $c=12 M r_+$ is the central charge. Different sequences $\{n_m\}$ correspond to different microstates of the Kerr black hole with same $(M,J)$.

The constraints (\ref{61}) equivalent to
\begin{equation}\label{62}
  \sum_m |m| n_m=\frac{c M^2}{2},\quad \sum_m m n_m=\frac{c a M}{2},\quad m\neq 0.
\end{equation}
The calculation of the Kerr black hole entropy then transforms into a mathematical problem: count the number of all different sequences $\{n_m\}$ that satisfy the constraints (\ref{62}). The constraints are the same as that for the BTZ black hole, so we can get the entropy for the Kerr black hole
\begin{equation}\label{66}\begin{split}
  N(M,J) \simeq \frac{1}{c (M^2+J) c (M^2-J)}\exp(2 \pi \sqrt{c \frac{M^2+J}{24}}+2 \pi \sqrt{c \frac{M^2-J}{24}}),\\
  S=\ln N(M,J)=  2 \pi M r_+-\ln [M^4 (r_+^2-a^2)^2]+\cdots,
\end{split}\end{equation}
which is just the Bekenstein-Hawking entropy formula with some low order corrections.

For extremal Kerr black hole $a=M$, the entropy is given by
\begin{equation}\label{67}
  S=\ln N(M,J)= \ln(\frac{1}{c a M} \exp(2\pi \sqrt{\frac{c a M}{12}}))=2\pi M^2-2\ln (M^2)+\cdots,
\end{equation}
which coincide the result of Ref.\cite{kerrfluff} for log-corrections.
\section{Conclusion}
In this paper, we analyze the boundary modes on the horizon of black holes with the method developed for topological insulator. Firstly the BTZ black hole is analysed, and the results are compatible with the previous works. Then we apply the same mathods to Kerr black holes. Some new results are obtained: dimensionless right- and left-temperature can be defined and have well behaviors both in Schwarzschild limit $a\rightarrow 0$ and extremal limit $a\rightarrow M$. Also a central charge $c=12 M r_+$ is associated with arbitrary Kerr black hole if we assume a dual CFT exist. We also identify the microstates of the Kerr black hole with the quantum states of the scalar field on the horizon. From this identification one can count the number of microstates of the Kerr black hole. The calculation of the Kerr black hole entropy transforms into a mathematical problem: count the number of all different sequences $\{n_m\}$ that satisfy some constraints. The final result gives the Bekenstein-Hawking entropy formula with some low order corrections.

Due to the compactness of the scalar field, one can get interesting results that the area of the black hole are quantized with an equally spaced spectrum both for BTZ black holes (\ref{29d}) and Kerr black holes (\ref{60a}). Those results consistent with the early works \cite{quans1,quans2}.

Based on the ``horizon fluff" proposal, the microstates of the extremal Kerr black hole are identified in Ref.\cite{kerrfluff}. The number of those microstates gives the Bekenstein-Hawking area law. This approach rely on the $U(1)$ Kac-Moody algebra and Virasoro algebra, which has close relation with Kerr/CFT correspondence. In our approach, we identify the microstates of Kerr black hole with the quantum states of the massless scalar field, and the algebra is just the commutation algebra for harmonic oscillator. Can one get the Virasoro algebra from our approach is under investigate. A related issue is the value of the central charge in entropy calculation formula (\ref{25e}) and (\ref{61}). We can not derive this value from first principle, since we do not have a conformal field theory and revelent Virasoro algebra. It is put by hand in order to get the right entropy. But it takes the same form for all Kerr black holes, which can be considered as a non-trivial check.

\acknowledgments
 This work is supported by Nanhu Scholars Program for Young Scholars of XYNU.
%\section{Appendix}
%\bibliography{bms5}

\end{document}